\documentclass[12pt,preprint]{aastex}
\IfFileExists{srcltx.sty}{\usepackage[active]{srcltx}}%

\def\gtap{\mathrel{ \rlap{\raise 0.511ex \hbox{$>$}}{\lower 0.511ex
   \hbox{$\sim$}}}} \def\ltap{\mathrel{ \rlap{\raise 0.511ex
   \hbox{$<$}}{\lower 0.511ex \hbox{$\sim$}}}} 
\newcommand{\beq}{\begin{equation}}

\newcommand{\eeq}{\end{equation}}
\newcommand{\bea}{\begin{eqnarray}}
\newcommand{\eea}{\end{eqnarray}}

\title{Thermal evolution of the primordial clouds in warm dark matter models 
with keV sterile neutrinos}

\author{Jaroslaw Stasielak\altaffilmark{1,2,3}, Peter L.
Biermann\altaffilmark{2,3,4},
and Alexander Kusenko\altaffilmark{5} }

\altaffiltext{1}{Institute of Physics, Jagiellonian University, ul. Reymonta 4,
30-059 Krak\'ow, Poland} 
\altaffiltext{2}{
Max-Planck Institute for Radioastronomy, Bonn, D-53121, Germany}
\altaffiltext{3}{
Department of Physics and Astronomy, University of Bonn, D-53121, Germany}
\altaffiltext{4}{
Department of Physics and Astronomy, University of Alabama, Tuscaloosa, 
AL 35487, USA }
\altaffiltext{5} {
Department of Physics and Astronomy, University of California, Los
Angeles, CA 90095-1547, USA }

\begin{document}



\begin{abstract}
We analyze the processes relevant for star formation in a model with dark
matter in the form of sterile neutrinos.   Sterile neutrino decays produce an
x-ray background radiation that has a two-fold effect on the collapsing clouds
of hydrogen.  First, the x-rays ionize the gas and cause an increase in the
fraction of molecular hydrogen, which makes it easier for the gas to cool and
to form stars.  Second, the same x-rays deposit a certain amount of heat,
which could, in principle, thwart the cooling of gas.  We find that, in all the
cases we have examined, the overall effect of sterile dark matter is to
facilitate the cooling of gas.  Hence, we conclude that dark matter in the form
of sterile neutrinos can help the early collapse of gas clouds and the
subsequent star formation. 
\end{abstract}


\section{Introduction}
\label{sec_introduction}

Numerous observations, from galaxy rotation curves to
gravitational lensing, to cosmic microwave background radiation, all point to
the existence of dark matter which is not made of ordinary atoms, but, rather,
of some new, yet undiscovered particles.  In particular, the abundance of dark
matter has been precisely determined recently by the
WMAP~\citep{spergel,spergel06}.  However, the nature of dark-matter particles
remains unknown.  One attractive candidate is a sterile neutrino with mass of
several keV and a small mixing with the ordinary neutrinos.  If such a particle
exists, it could be produced in the early universe with the right abundance to
be dark matter either from neutrino
oscillations~\citep{Dodelson,shi,abaz01a,abaz01b,abaz05,Dolgov02}, or from some
other processes, for example from inflaton decay~\citep{Shaposhnikov:2006xi}.  
The same particle can explain the observed velocities of pulsars because its
emission from a cooling neutron star would be anisotropic, hence providing the
neutron star with a recoil velocity of a sufficient magnitude
\citep{Kusenko97,b11,Kusenko98,fuller,barkovich,2005hep.ph....3113B,Kusenko04}.
In addition, the neutrino kick can enhance the convection during the first
second of the supernova explosion, which can increase the energy of the
supernova shock and bring the supernova calculations in better agreement with
the observations \citep{fryer}. Sterile neutrinos can also help the
formation of supermassive black holes in the early universe \citep{mun,bb}. 

Although the dark-matter sterile neutrinos have interactions that are too
weak to be discovered in the laboratory,  some other sterile neutrinos may
exist. Most models of neutrino masses introduce
sterile (or right-handed) neutrinos to generate masses of the ordinary
neutrinos via the seesaw mechanism~\citep{seesaw1,seesaw2,seesaw3}.  The name
{\em sterile} was coined by~\cite{pontecorvo}.  Many seesaw models assume that
sterile neutrinos have very large masses, which makes them unobservable. 
However, a number of theoretical models predict the existence of sterile
neutrinos at or below the
electroweak scale~\citep{smirnov,shrock,asakaa,deGouvea}. There is an
indication that some light sterile mass eigenstates may
cause the neutrino flavor transformation seen in the Los Alamos Liquid
Scintillator Neutrino Detector (LSND) experiment \citep{b40,sorel,deGouvea}.  
The results of LSND are being tested by the MiniBooNE experiment. 
One can expect that there are at least three light sterile
neutrinos \citep{asakab}, in which case the neutrino oscillations can explain
the baryon asymmetry of the universe \citep{akhmedov,asakaa}.

Although dark-matter sterile neutrinos are stable on cosmological time scales,
they nevertheless decay. The mass matrix is assumed to have the seesaw form
with the Majorana mass term of the order of several keV and with a much smaller
Dirac mass term.  The diagonalization of such a mass matrix yields the light
mostly active Majorana mass eigenstates and the heavy sterile Majorana
eigenstates, which can decay into the light ones.  The dominant decay mode,
into three light neutrinos,
is "invisible" because the daughter neutrinos are beyond the detection
capabilities of today's experiments.  The most prominent "visible" mode is
decay into one active neutrino and one photon, $\nu_s \rightarrow \nu_a
\gamma$.  Assuming two-neutrino mixing for simplicity, one can express the
inverse width of such a decay as 
\begin{equation}
\tau \equiv \Gamma_{\nu_s \rightarrow \nu_a
\gamma}^{-1}= 1.3 \times 10^{26} \rm{ s } \left(\frac{7\ \rm keV}{m_{s}
}\right)^{5} \left(\frac{0.8 \times 10^{-9}}{\sin^{2}
\theta}\right),  
\label{meanlife}
\end{equation}
where $m_s$ is the mass and $\theta $ is the mixing angle.  For mass units we
set $c=1$.

\begin{figure}
\plotone{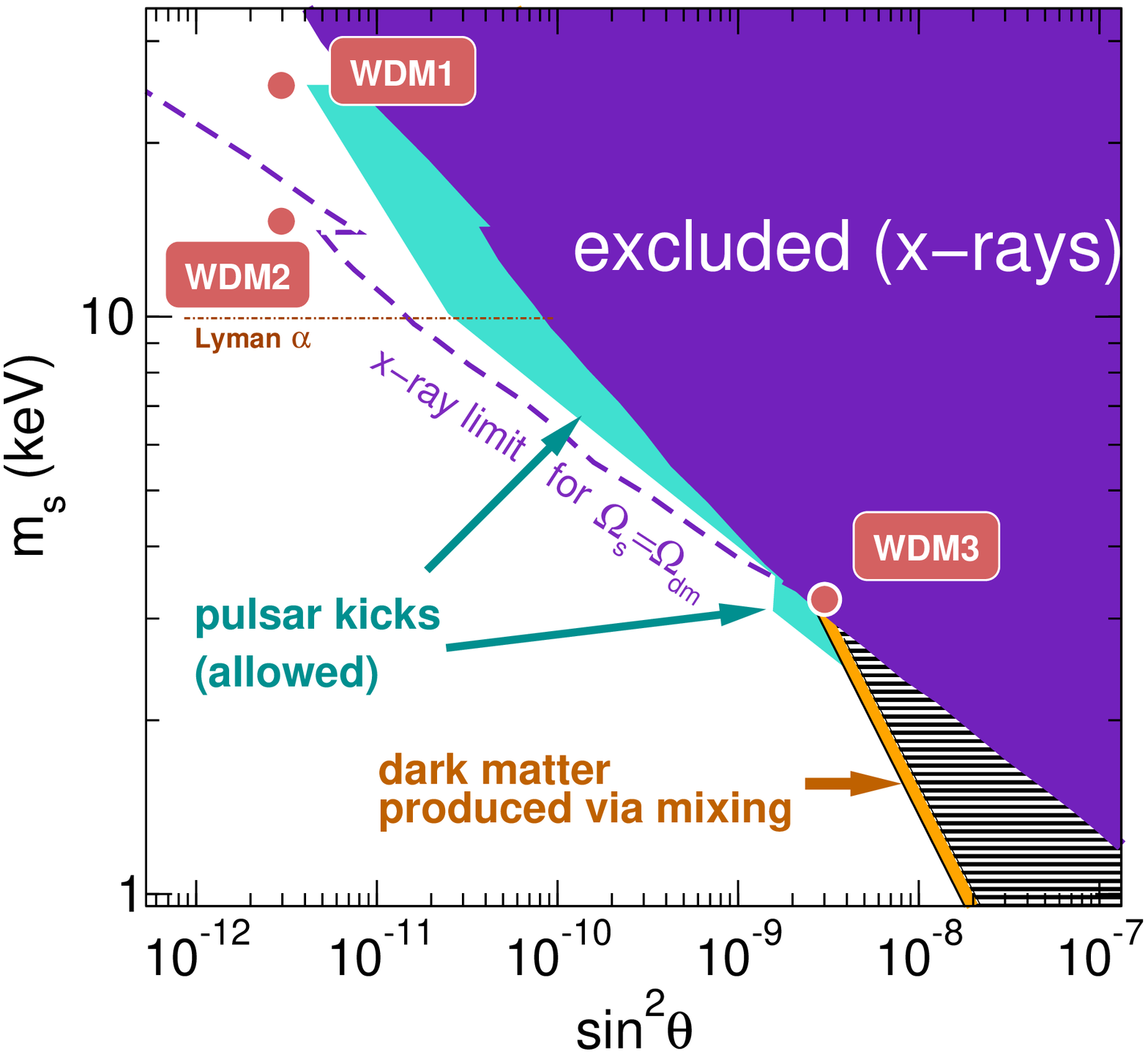}
\caption{
The parameter range favored by the pulsar kicks, the current limits from x-ray
observations (see text for discussion), and the benchmark points chosen for
model calculations. 
}
\label{fig:models}
\end{figure}

Since this is a two-body decay, the photon energy is half the mass of
the sterile neutrino.  The monochromatic line from dark matter decays can, in
principle, be observed by x-ray telescopes.  No such observation has been
reported, and some important limits have been derived on the allowed masses and
mixing angles~\citep{abaz01a,abaz01b, 
Boyarsky,2006hep.ph....1098B,2006astro.ph..3368B,2006astro.ph..3660B,
Riemer-Sorensen:2006fh,2006PhRvD..73f3513A,Watson:2006qb}.   These constraints
are based on different astrophysical objects, from Virgo and Coma clusters, to
Large Magellanic Clouds, to the Milky Way halo and its components, etc.  There
are different uncertainties in modeling the dark matter populations in these
objects. Different groups have also used very different methodologies in
deriving these bounds: from a conservative assumption that the dark-matter line
should not exceed the signal, to more ambitious approaches that involved
modeling the signal or merely fitting it with a smooth curve and requiring that
the line-shaped addition not affect the quality of the fit.  The x-ray limits
are usually shown based on the assumption that the sterile neutrinos
constitute the entire dark matter ($\Omega_s=\Omega_{dm}$, where $s$ and $dm$
refer to the sterile neutrinos and the dark matter, respectively). However,
the x-ray observations exclude certain masses and mixing angles even if the
sterile neutrinos make up only a fraction of dark matter.  The production via
mixing \citep{Dodelson} provides the lowest possible abundance of sterile
neutrinos. In Fig.~\ref{fig:models} we show both the bounds based on the
assumption $\Omega_s=\Omega_{dm}$ and the model-independent exclusion
region~\citep{Kusenko06} based on the production only via the \cite{Dodelson}
mechanism.  \cite{Kusenko06} used 
the analytical fit to the numerical calculation of \cite{abaz05}, and 
re-scaled the x-ray flux in proportion to the amount of relic neutrinos
produced for a given mass and mixing angle.

A  different set of bounds comes from the observations of the
Lyman-alpha forest~\citep{viel1,viel2,seljak}, which limit the sterile
neutrino mass from below. Based on the high-redshift data from SDSS and some 
modeling of gas dynamics, one can set a limit as strong as
14~keV~\citep{seljak}. However, the high-redshift data have systematic errors
that are not well understood~\citep{jena}, and more conservative approaches,
based on the relatively low-redshift data, have led to some less stringent
bounds~\citep{viel1}.  More recently,
\cite{viel2} have reanalyzed the high-redshift data and arrived at the bound
$m_s>10$~keV.  The mass bound as quoted depends on the production mechanism
in the early universe.  

The Lyman-alpha observations constrain the free-streaming lengths of dark
matter particles, not their masses.  For each cosmological production
mechanism, the relation between the free-streaming length and the mass is
different~\citep{Hansen_et_al}.  For example, the bound $m_s
>10$~keV~\citep{viel2} applies to the
production model due to~\cite{Dodelson}.  Even within a given cosmological
scenario, there are uncertainties in the production rates of neutrinos for any
given mass and mixing angle~\citep{Asaka:2006rw}.  These uncertainties may
further affect the interpretation of the Lyman-alpha bounds in terms of the
sterile neutrino mass. The x-ray bounds depend on a combination of mass and
mixing angle as in eq.~(\ref{meanlife}), regardless of one's cosmological
assumptions, as long as all the dark matter is comprised of sterile neutrinos.
However, the x-ray bounds are subject to uncertainties in modeling the dark
matter populations of different objects. 

If the sterile neutrinos make up only a part of dark matter, 
none of the above bounds apply.  Also, if inflation ended with a low reheating 
temperature, the bounds are significantly weaker~\citep{b22}.

It should also be mentioned that the Lyman-alpha bounds appear to contradict
the observations of dwarf  spheroidal
galaxies~\citep{Wilkinson,gilmore_06,strigari},
which  suggest that dark matter is warm and which would favor the 1--5~keV mass
range for sterile neutrinos. There are several inconsistencies between the
predictions of N-body simulations of cold dark matter (CDM) and the
observations
\citep{willman,bode,Peebles,Dalcanton,bosch,zentner,abaz05b,Dolgov01,
b38,b39,halo_shape,halo_shape2,halo_shape_cdm}.
Each of these problems may find a separate independent solution.   Perhaps, a
better understanding of CDM on small scales will resolve these discrepancies.  
It is true, however, that warm dark matter in the form of sterile neutrinos
is free from all these small-scale problems altogether, while on large scales
WDM fits the data as well as CDM.

Several independent considerations, from the pulsar kicks to small-scale
behavior of dark matter, favor sterile neutrinos as the dark matter candidate.
It is important, therefore, to examine the effects of sterile-neutrino dark
matter on star formation and reionization of the universe. \citet{yoshida03}
have attempted to study such effects for some kind of warm dark matter, but
they have used a linear matter power spectrum and assumed that the dark matter
particles have a thermal distribution, which is not the case for sterile
neutrinos \citep{abaz05}. Even more importantly, they have not considered the
effects of sterile neutrino decays and the resulting x-ray background. It was
shown by \citet{biermann06} that the x-rays produced by the decays of relic
sterile neutrinos can speed up the formation of molecular hydrogen, which can
help the cooling of the gas and the star formation, which can lead, in turn, to
a reionization of the universe at the redshift consistent with the WMAP
results.  

More specifically, the WMAP (three years) measurement of the reionization
redshift $z_r=10.9^{+2.7}_{-2.3}$ \citep{spergel06} has posed a challenge to
theories of star formation. On the one hand, stars have to form early enough to
reionize gas at redshift~11. On the other hand, the spectra of bright distant
quasars imply that reionization must be completed by redshift~6.  Stars form in
clouds of hydrogen which collapse at different times, depending on their sizes:
the small clouds collapse first, while the large ones collapse last.  If the
big clouds must collapse by redshift~6, then the small halos must  undergo the
collapse much earlier.  It appears that the star formation in these small halos
would have occurred at high redshift, when the gas density was very high, and
it would have resulted in an excessive Thompson
optical depth \citep{haiman_bryan}. To be consistent with WMAP, the efficiency
for the production of ionizing photons in minihalos must have been at least an
order of magnitude lower than expected~\citep{haiman_bryan}. One solution is to
suppress the star formation rate in small halos by some dynamical feedback
mechanism. The suppression required is by at least an order of magnitude.  

Alternatively, one can consider warm dark matter, in which case the number of
small clouds is strongly suppressed.  However, for some WDM candidates (for
example, the gravitino), the lack of power on small scales in the spectrum of
density perturbations leads to an unacceptable delay in the onset of star
formation \citep{yoshida03}. This is not the case for sterile neutrinos, which
decay slowly during the "dark ages" preceding the birth of stars.  The decay
photons could ionize gas, and, although this ionization is too weak to
affect the WMAP measurements directly~\citep{mapelli1,mapelli2}, the increase
in the ionization fraction could cause a rapid production of molecular hydrogen
at redshifts as high as 100~\citep{biermann06}. Molecular hydrogen is the
essential cooling agent that allows the gas to cool and collapse to form the
first stars.  

\citet{biermann06} have not considered the effects of gas
heating associated with the decays, which may, in some cases,  thwart the
effects of ionization.   The increase in the fraction of molecular hydrogen can
help cooling the gas, but if the heating from sterile neutrino decays is
significant, it can overcome the radiative cooling and stymie the collapse and
star formation.   In this paper we will examine thermal evolution of gas clouds
in detail, taking into account both effects of the sterile neutrino decays,
namely the ionization and the heating of gas. 

The paper is organized as follows. In Section \ref{sec_H2} we will discuss the
role of sterile neutrino decays and the production of molecular hydrogen in the
cooling of the primordial gas clouds. Next, in Sections 3-6 we will briefly
describe the top-hat overdensity simulation. The results are shown in
Section~\ref{sec_results}.   

\section{H$_{2}$ cooling and sterile neutrinos decays}
\label{sec_H2}

The existence of the first objects is a direct consequence of the growth
of the primordial density fluctuations. At the beginning, there are linear
density perturbations which expand with the overall Hubble flow. Subsequently,
these perturbations can grow and form primordial clouds. Clouds
with enough density contrast decouple from this flow and start to collapse.
The kinetic energy of the infalling gas is dissipated through shocks and the
cloud becomes pressure supported. The subsequent evolution of the cloud is
determined by its ability to cool sufficiently fast. Clouds that do not
cool fast enough stay in a pressure-supported state and do not form 
stars. The existence of the efficient cooling mechanism is necessary to
continue the collapse of the cloud, its subsequent fragmentation and star
formation.

In the absence of metals, the most important cooling mechanism is cooling by
hydrogen atoms and molecules.  This mechanism involves the collisional
excitation of H or H$_{2}$, subsequent spontaneous de-excitation, and photon
emission. The emitted photons can escape and take away energy, hence reducing 
the kinetic energy of the collapsing cloud. Cooling by H dominates for
temperatures above $10^{4}$~K, while the H$_{2}$ cooling is more important at
lower temperatures. The temperature of the primordial cloud can be higher than
$10^{4}$ K only during a short period of virialization of the baryonic gas.
Therefore, H$_{2}$ cooling is the most important cooling mechanism during
almost entire evolution of the halos, and it is crucial for the star
formation.

In primordial gas clouds, H$_{2}$ molecules can form mainly through the coupled
reactions
\begin{eqnarray}
	e^{-} + H& \rightarrow &H^{-} + \gamma  \\
	H^{-}+H & \rightarrow & H_{2} + e^{-}  \textrm{,}
\end{eqnarray}
or
\begin{eqnarray}
	H^{+} + H& \rightarrow &H_{2}^{+} + \gamma  \\
	H_{2}^{+}+H & \rightarrow & H_{2} + H^{+}  \textrm{,}
\end{eqnarray}
in which $e^{-}$ and $H^{+}$ act only as a catalyst. X-ray radiation can
increase the production of the H$_{2}$ by enhancement of the ionization
fraction.

As discussed above, sterile neutrinos with the mass of several keV and a small
mixing are stable on cosmological time scales. Nevertheless some of them can
decay.  The decay channel important for us is
\begin{equation}
	\nu_{s} \rightarrow \nu_{a} +\gamma \textrm{.}
\end{equation}
Its inverse width is given in equation (\ref{meanlife}). 
The energy of the photon is equal to
\begin{equation}
	E_{0}=\frac{\left(m_{ s}^{2}+m_{ a}^{2}\right)}{2m_{ s}} \approx
\frac{m_{ s}}{2},
\end{equation}
where $m_{ a}$ is the mass of the active neutrino. X-ray photons from the
sterile neutrinos decays can increase the ionization fraction and the H$_{2}$
production. This can subsequently lead to a speed-up of the gas cooling and
star formation. 

\section{Density evolution}
\label{sec_density}

We will follow the evolution of the baryonic top-hat overdensity, the gas
temperature and its H$_{2}$ and $e^{-}$ fraction.  Our goal is to juxtapose
the evolution of the gas temperature in the primordial clouds in the 
$\Lambda$CDM model and the WDM model with keV sterile neutrinos. 
We assume that the dark matter is smoothly distributed in the region we are
considering.

We will use the parameters from the best fit to the three years WMAP data 
\citep{spergel06}, namely $H_{0}=100h$ km $s^{-1}$ $Mpc^{-3}$, $h=0.73$,
$\Omega_{0}=\Omega_{dm}+\Omega_{b}=0.238$, $\Omega_{b}=0.042$,
$\Omega_{dm}=\Omega_{0}-\Omega_{b}=0.196$, $\Omega_{\Lambda}=0.762$,
$\Omega_{tot}=\Omega_{dm}+\Omega_{b}+\Omega_{\Lambda}=1$.

Let us take some spherical region with the top-hat overdensity $\delta$. It is
well known that such an overdensity will evolve according to the following
equation \citep{Padmanabhan}
\begin{equation}
	1+\delta=\frac{9}{2}\frac{\left(\alpha-\sin
\alpha\right)^{2}}{\left(1-\cos \alpha\right)^{3}} \textrm{,} \label{th1}
\end{equation}
where the parameter $\alpha$ is related to the redshift $z$ and redshift of
virialization $z_{vir}$ through
\begin{equation}
	\frac{1+z_{vir}}{1+z}=\left(\frac{\alpha- \sin \alpha}{2
\pi}\right)^{2/3}. \label{th2}
\end{equation}

Further evolution depends on the type of the matter in the overdense 
region. If it is the dark matter then after virialization its density remains 
constant forever. The situation is different in the case of baryons. If there
exists an efficient cooling mechanism, then the density gradually
increases. Otherwise, the density remains constant, and there is no star
formation in the halo.

In reality an overdense region consists of two components namely, baryonic gas
and dark matter. Baryonic density fluctuations are suppressed before
recombination. In contrast, the dark matter density fluctuations can start to
grow well before this period, so its initial overdensity is larger than that
of the baryons.  Dark matter forms the potential wells into which the baryonic
matter falls.  Overall, the baryon density follows the dark matter
density in the early universe.  Therefore, one can consider the primordial
fluctuations as the halos consisting of baryonic and dark matter, whose 
overall overdensity follow the top-hat solution (eqs. \ref{th1} and \ref{th2}).
Of course, this assumption breaks down at the later stages of the halo
evolution, e.g. after the virialization of the dark matter.

Following \cite{tegmark97}, let us assume that the evolution of the
baryonic and dark matter density is described by equations (\ref{th1}) and
(\ref{th2}) until $\alpha$ reaches $3\pi/2$. This occurs at  the redshift
\begin{equation}
	z_{3\pi/2}=1.06555\left(1+z_{vir}\right)-1 \textrm{.} \label{z3pi}
\end{equation} 
After this redshift, we assume that the density of the halo $\rho_{b+dm}$ stays
constant e.g. it is equal to the value
\begin{equation}
	\rho_{b+dm}=18\pi^{2}\Omega_{0}\rho_{0}\left(1+z_{vir}\right)^{3}
\textrm{,} \label{densvir}
\end{equation}
where $\rho_{0}$ is the critical density of the universe.

We note that the overdensity $\delta$ in equation (\ref{th1}) is
given by
\begin{equation}
	\delta=\frac{\rho_{b+dm}}{\bar{\rho}}-1 \textrm{,}
\end{equation}
where $\bar{\rho}$ is the combined average density of dark and baryonic matter.
 
We have taken the ratio of the baryonic to the dark matter mass in the halo to
be the standard cosmological ratio of $\Omega_{b}/\Omega_{dm}\approx0.214$. The
density of the baryonic matter can be expressed as 
\begin{equation}
	\rho_{b}=\rho_{b+dm}\frac{\Omega_{b}}{\Omega_{0}} \textrm{.}
\end{equation}

\section{Temperature evolution}
\label{sec_temperature}

The gas temperature evolution is governed by the equation of energy
conservation
\begin{equation}
	\frac{du}{dt}=\frac{p}{\rho_{b}^{2}}\frac{d\rho_{b}}{dt}-\frac{\Lambda}{
 \rho_{b}} \textrm{,} \label{u1}
\end{equation}
where $p$, $u$ and $\Lambda$ are the pressure, the internal energy per unit
mass and the cooling/heating function respectively. We use the equation of
state of perfect gas
\begin{equation}
	p=\left(\gamma-1\right)\rho_{b} u \textrm{,} \label{p}
\end{equation}
where $\gamma$ is the adiabatic index.

The internal energy per unit mass can be described as
\begin{equation}
	u=\frac{1}{\gamma-1}\frac{kT}{\mu m_{H}} \textrm{,} \label{u}
\end{equation}
where $k$ is the Boltzmann constant, $T$ is the gas temperature, $\mu$ denotes
the  molecular weight, and $m_{H}$ is the mass of the hydrogen atom.
Introducing the number density of non dark matter particles $n_{p}$ we can
write $\rho_{b}$ as
\begin{equation}
	\rho_{b}=\mu m_{H}n_{p} \textrm{.} \label{p2}
\end{equation}

If we use eqs. (\ref{p}), (\ref{u}) and (\ref{p2}) than we can cast eq.
(\ref{u1}) in the form
\begin{equation}
\frac{dT}{dt}=\left(\gamma-1\right)\left(\frac{T}{n_{p}}\frac{dn_{p}}{dt}-\frac
{\Lambda}{n_{p}k}\right)+\gamma\frac{T}{\mu}\frac{d\mu}{dt}+\frac{T}{
\left(\gamma-1\right)}\frac{d\gamma}{dt} \textrm{.} \label{tt}
\end{equation}

The time $t$ can be translated to the redshift $z$ in the following way
\begin{equation}	
\frac{dt}{dz}=-\frac{1}{H_{0}\left(1+z\right)\sqrt{\Omega_{\Lambda}+\Omega_{0}
\left(1+z\right)^{3}}} \textrm{.} \label{hubble}
\end{equation}
Using eqs. (\ref{tt}) and (\ref{hubble}),  we obtain 
\begin{equation}
\frac{dT}{dz}=\left(\gamma-1\right)\frac{T}{n_{p}}\frac{dn_{p}}{dz}+\gamma\frac
{T}{\mu}\frac{d\mu}{dz}+\frac{T}{\left(\gamma-1\right)}\frac{d\gamma}{dz}+ 
	\frac{\left(\gamma-1\right)\Lambda}{n_{p}k
H_{0}\left(1+z\right)\sqrt{\Omega_{\Lambda}+\Omega_{0}\left(1+z\right)^{3}}}
\textrm{.} \label{tz}
\end{equation}
To calculate the gas temperature of the primordial cloud we have to integrate
this equation.

One can assume that the gas temperature evolution is governed by eq.
(\ref{tz}) except for the redshift range between $z_{3\pi/2}$ and $z_{vir}$. If
we simply integrate eq.(\ref{tz}) to the redshift $z_{vir}$, than the gas
temperature will be much lower than the virial temperature $T_{vir}$, at least
for clouds with mass greater than some critical value~\citep{kitayama01}
\begin{equation}	
T_{vir}= 9.09\times10^{3}\textrm{ K
}\left(\frac{\mu_{vir}}{0.59}\right)\left(\frac{M_{b+dm}}{10^{9}h^{-1}M_{\sun}}
\right)^{2/3}  
 \left(\frac{\Delta_{c} \left( z_{vir}\right)}{18 \pi^2}\right)^{1/3} 
\left(1+z_{vir}\right) \textrm{,} \label{virial}
\end{equation}
where $\mu_{vir}$, $M_{b+dm}$ and $\Delta_{c} \left(z_{vir}\right)$ are
respectively the mean molecular weight during virialization, the halo mass and
the overdensity of the halo at virialization given by \citep{bryan98},
which is very well approximated by $18\pi^2$ for redshifts $z_{vir} \leq 100$.
Therefore we have to take into account shocks and increase the gas temperature
to the virial value. It is reasonable to assume that the gas temperature
evolution is linear between the redshift $z_{3\pi/2}$ and $z_{vir}$ and is
described by the equation
\begin{equation}
	\frac{dT}{dz}=\frac{T_{vir}-T_{3\pi/2}}{z_{vir}-z_{3\pi/2}} \textrm{,}
\label{telin}
\end{equation}
where $T_{3\pi/2}$ is the temperature at the redshift $z_{3\pi/2}$.

Let us consider the following five species: H, H$^{+}$, ${\rm H}_{2}$,
$e^{-}$, and $\rm He$ with the mass fraction $Y=0.244$ \citep{izotov}.
Let us denote by $n_{i}$ ($i=$H, H$^{+}$, H$_{2}$, $e^{-}$ and He) the number
density of baryonic component $i$. We assume that $n_{{\rm H}^{+}}=n_{e}$ e.g.
there is no
He or H$_{2}$ ionization. The number density of hydrogen can be written as
\begin{equation}
	n=n_{H}+n_{H^{+}}+2n_{H_{2}} \textrm{.}
\end{equation}
If we introduce the number fraction of component "$i$"   
\begin{equation}
	x_{i}=\frac{n_{i}}{n} \textrm{,}
\end{equation}
then the particle number density $n_{p}$ and the baryon density $\rho_{b}$
are equal to
\begin{equation}
	n_{p}=n\left(1+\frac{Y}{4X}-x_{H_{2}}+x_{e}\right)  \label{nbb} 
\end{equation}
and
\begin{equation}
	\rho_{b}=\frac{n m_{H}}{X}	\label{rob} \textrm{,}
\end{equation}
where $X=1-Y=0.756$ is the hydrogen mass fraction.

Using eqs. (\ref{p2}), (\ref{nbb}) and (\ref{rob}) we can get
\begin{equation}
	\mu=\frac{1}{X\left(1+x_{e}-x_{H_{2}}\right)+0.25Y} \textrm{.}
\end{equation}
At the virialization we have typically $x_{H_{2}}\approx0$ and $x_{e}\approx
1$,
so one can take the mean molecular weight at this time to be
$\mu_{vir}=1/\left(2X+0.25Y \right)\approx0.63$. We can express $\gamma$ as
\begin{equation}
	\gamma=1+\left(\sum_{i=1}^{5}n_{i}\right)/\left(\sum^{5}_{i=1}
	\frac{n_{i}}{\gamma_{i}-1}\right) \textrm{,}
\end{equation}
where $\gamma_{H}=\gamma_{H^{+}}=\gamma_{e}=\gamma_{He}=5/3$ and
$\gamma_{H_{2}}=7/5$.

\section{Chemical reactions and cooling/heating function}
\label{sec_chemical}

To calculate the cooling/heating function $\Lambda$ in eq. (\ref{tz}) we have
to take into account all of the relevant chemical and thermal processes and
follow the number density evolution of different baryonic components. All 
chemical and thermal processes included in our calculation are summarized in
Tables \ref{tab:reaction} and \ref{tab:cooling}. The evolution of $x_{e}$ and
$x_{H^{+}}$ is described by equations
\begin{equation}
	\frac{d
x_{e}}{dt}=\left(A\left(z\right)+k_{12}+k_{14}n_{e}\right)x_{H}-k_{13}x_{e}^{2}
n \label{xe}
\end{equation}
\begin{equation}
	\frac{dx_{H^{+}}}{dt}=\frac{dx_{e}}{dt} \textrm{,}
\end{equation}
where $A\left(z\right)$ is the production rate of free electrons by the sterile
neutrino decays which is calculated in the appendix. In eq. (\ref{xe}) Only
the ionization and recombination of hydrogen are included. All the
rates $k_{i}$ are given in Table \ref{tab:reaction}. 
 
We note in passing that neglecting He and H$_{2}$ ionization renders the
system with fewer free electrons, so the production of H$_{2}$ is slower.
In addition, cooling by H$_{2}^{+}$ is stronger than by H$_{2}$ (see last
figure in \citet{galli}). Therefore, the cooling we get in this approximation
is somewhat less efficient than it should be in reality. 

The molecular hydrogen number fraction can be calculated by integration of the
following equation
\begin{equation}
	\frac{dx_{H_{2}}}{dt}= n_{H}\left(   x_{e} k_{m}-k_{7}x_{H_{2}}
\right)-
	\left(k_{8}x_{H_{2}}+k_{9}x_{e}\right)n_{H_{2}}
	-\left(k_{10}+k_{11}\right)x_{H_{2}} \rm{,}
\end{equation}
where
\begin{equation}
	k_{m}=\frac{k_{1}k_{2}}{k_{2}+k_{3}/n_{H}}+\frac{k_{4}k_{5}}{k_{5}+k_{6}
 /n_{H}} \label{km}
\end{equation}
is the molecular hydrogen formation rate via $H^{-}$ and H$_{2}^{+}$ channel
\citep{tegmark97}. To get eq. (\ref{km}) we have to make the assumption that
after formation of $H^{-}$ and $H^{+}_{2}$ they are used almost instantaneously
in the next reactions (reactions number 2 and 3 or 5 and 6 respectively, see
Table \ref{tab:reaction}). This is a very good approximation, because
in reality the $H^{-}$ and $H^{+}_{2}$ fractions always remain small. 

The cooling/heating function $\Lambda$ is given in the appendix (eq.
\ref{coolfunc}). It includes contributions from the following processes:
\begin{enumerate}
	\item excitation of H by collisions with $e^{-}$
	\item recombination to H
	\item ionization of H by collisions with $e^{-}$
	\item photo-ionization of H by the CMB radiation
	\item excitation of H$_{2}$ by collisions with H and H$_{2}$
	\item H$_{2}$ formation cooling
	\item dissociation of H$_{2}$ by collisions with H, H$_{2}$ and
$e^{-}$
	\item photo-dissociation of H$_{2}$ by CMB photons
	\item Compton cooling
	\item bremsstrahlung 
	\item photo-ionization of H by the photons from the sterile neutrino
decays and H ionization by the secondary electrons produced due to these
photons
\end{enumerate}
We have not included the effect of the photo-dissociation of H$_{2}$ by the
photons from the sterile neutrino decays. This is because the flux
of these photons is small in comparison with the CMB radiation in the 10 - 30
eV range, which is the energy range relevant to the photo-dissociation of
H$_{2}$. 

The formulae for the contributions to the $\Lambda$ function from all of these
processes are given in Table \ref{tab:cooling}. We have assumed the
ortho-H$_{2}$ to para-H$_{2}$ ratio to be 3 : 1. 

\section{Initial conditions and numerical procedure}
\label{sec_init}

We  start our calculations at the redshift of recombination $z_{rec}=1013$,
taking as initial conditions $x_{e}=1$, $x_{H_{2}}=0$ and assuming that the gas
temperature is equal to the CMB temperature. Evolution of the halo is
insensitive for the initial value of $x_{H_{2}}$.

We follow the evolution of the density, eqs. (\ref{th1}) and (\ref{th2}), and
the temperature, eq.~(\ref{tz}) with the cooling/heating function given in
eq.~(\ref{coolfunc}) untill the redshift is equal to $z_{3\pi/2}$,
\textit{cf.}~eq.~(\ref{z3pi}). From 
that time we assume that the density of the halo stays constant and is equal to
its virial value (\ref{densvir}). We assume that the gas starts to be
shock
heated in this period and reaches its virial temperature at $z_{vir}$. The
temperature of the gas is assumed to change linearly with the redshift between
$z_{3\pi/2}$ and $z_{vir}$, eq.~(\ref{telin}). For the redshift lower than
$z_{vir}$ the gas temperature is calculated once again by integration of 
eq.~(\ref{tz}).

Following \citet{tegmark97}, we use the following 
criterion for the successful collapse:  
\begin{equation}
	T\left( \eta z_{vir} \right) \leq \eta T\left( z_{vir} \right), 
\end{equation}
where $\eta=0.75$ \citep{tegmark97}.

\section{Results}

\label{sec_results}

The results of our analysis depend, of course, on the parameters we choose for
the dark-matter sterile neutrinos.   The x-ray flux, and the resulting
increase
in the ionization fraction depend on the combination ($m_s^5 \sin^2\theta$). 
The power spectrum on small scales, which determines the presence or absence of
small clouds, depends on the free-streaming length, which is set by the mass
(in every given cosmological scenario).   

Based in part on the existing limits, we have chosen four illustrative
benchmark cases to illustrate the effects of ionization and heating due to
sterile neutrino decays on the cloud collapse.

We have compared the following four models:

\begin{enumerate}
	\item Cold dark matter with no additional radiation besides CMB (CDM).
	
	\item Warm dark matter (WDM) with the sterile neutrino mass
$m_{s} =25$ keV and mixing angle $\sin^2 \theta =3 \times 10^{-12}$ (WDM1).

\item WDM  with $m_{s} =15$ keV and $\sin^2 \theta =3 \times 10^{-12}$
(WDM2).

\item WDM with $m_{s} =3.3$ keV and $\sin^2 \theta =3 \times 10^{-9}$
(WDM3).
\end{enumerate}
The values of parameters chosen for these models are shown in
Fig.\ref{fig:models}.  

Some comments are in order.   The choice of WDM1 is obvious because it
satisfies all the constraints, including the x-ray bounds and the Lyman-alpha
bounds.  The masses as large as $m_{s} =25$ keV cannot be constrained by
either Chandra or XMM, and the best constraint comes from HEAO, which is weaker
because of its lower energy resolution
\citep{Boyarsky,2006hep.ph....1098B,2006astro.ph..3368B,2006astro.ph..3660B}.
Also, WDM1 corresponds to a very small free-streaming length and, on large
scales, it creates structure indistinguishable from CDM.  The only difference
between WDM1 and CDM is  the additional x-ray flux.   Therefore, the comparison
of CDM and WDM1 cases provides a clean test of the effects of the increased
ionization fraction \citep{biermann06}.   Also, WDM1 provides a realistic set of
parameters consistent with all the bounds, as well as the pulsar kick mechanism
due to resonant active-sterile neutrino conversions \citep{Kusenko97}.  

WDM2 is also consistent with all the existing limits, including the most
ambitious Lyman-alpha bounds~\citep{seljak}.  (We remind the reader that 
\citet{viel2} have obtained a 50\% weaker limit using the same
data.)  The x-ray background is lower in this case than in the case of WDM1.  

Finally, WDM3 corresponds to masses that would be ruled out by the "strong"
Lyman-alpha bounds obtained by~\cite{seljak} and by~\cite{viel2}. However, 
one has to remember that the relation between the masses and the free-streaming
length depends on the production mechanism of sterile neutrinos.  For
example, the free-streaming length in the~\cite{shi} production mechanism is
different from the~\cite{Dodelson} mechanism.  Even within the
latter framework, additional entropy production can redshift the sterile
neutrino momenta and eliminate the Lyman-alpha bound~\citep{Asaka:2006ek}.  If
dark matter  is produced from a coupling between the sterile neutrinos and the
inflaton~\citep{Shaposhnikov:2006xi}, the
free-streaming length is, again, different.  WDM3 requires redshifting of the
neutrino momenta by a factor $\sim 3$, as compared to the thermal distribution.
This can be achieved, for example, by entropy production of the order of
$10^2$, which is comparable to the effect of change in the number of degrees
of freedom from the electroweak scale to zero temperature~\citep{Kusenko06}. 
Hence, WDM3 should not be considered an unrealistic scenario. In addition, WDM3
is of interest because it is consistent with the pulsar kick mechanism via
off-resonant active-sterile neutrino oscillations~\citep{fuller}, even if this
sterile neutrino is not the dominant component of dark matter.

\begin{figure}
\plotone{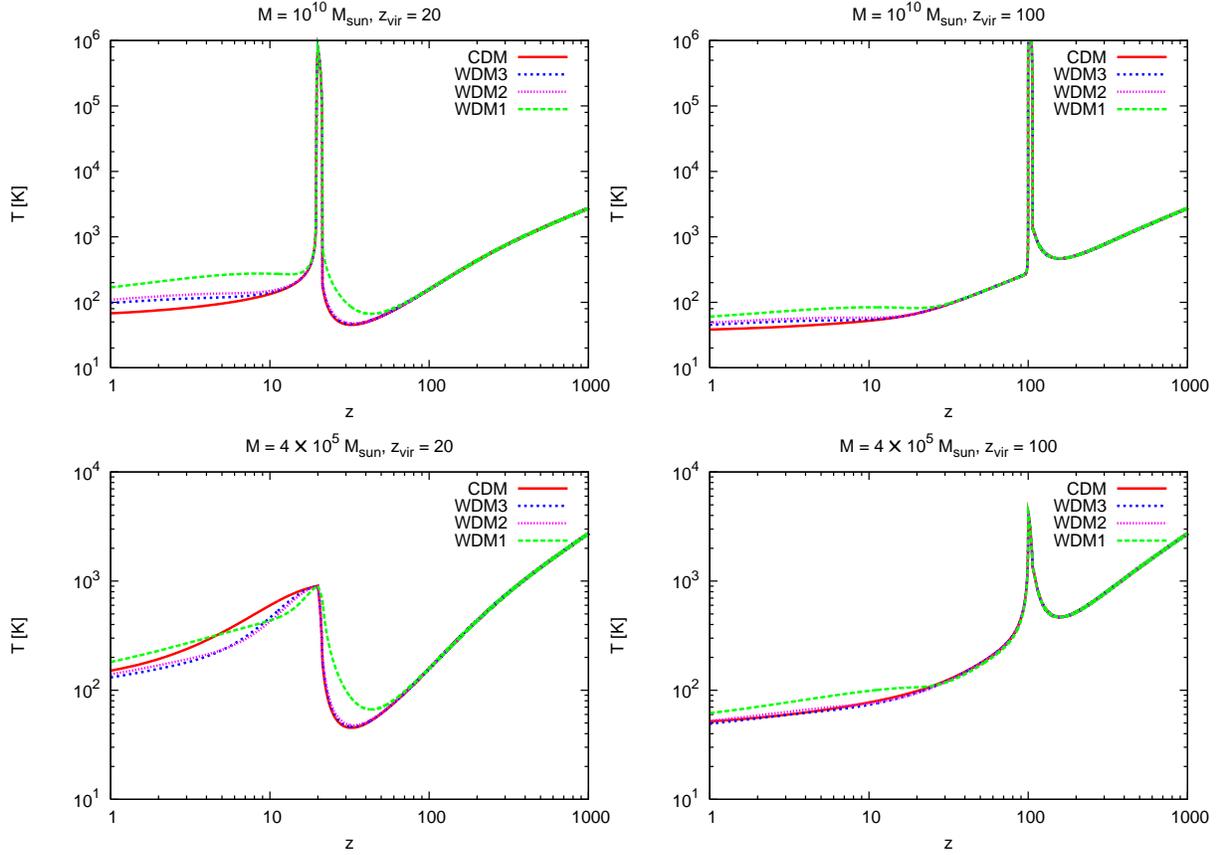}
\caption{Evolution of temperature with redshift for different models: $m_{s}
=25$~keV and $\sin^2 \theta=3 \times 10^{-12}$ (WDM1), $m_{s} =15$~keV and
$\sin^2 \theta=3 \times 10^{-12}$ (WDM2), $m_{s} =3.3$~keV and $\sin^2
\theta=3 \times 10^{-9}$ (WDM3). $M$ is the cloud mass and $z_{vir}$ is the 
redshift of virialization.}
\label{fig:T}
\end{figure}

\begin{figure}
\plotone{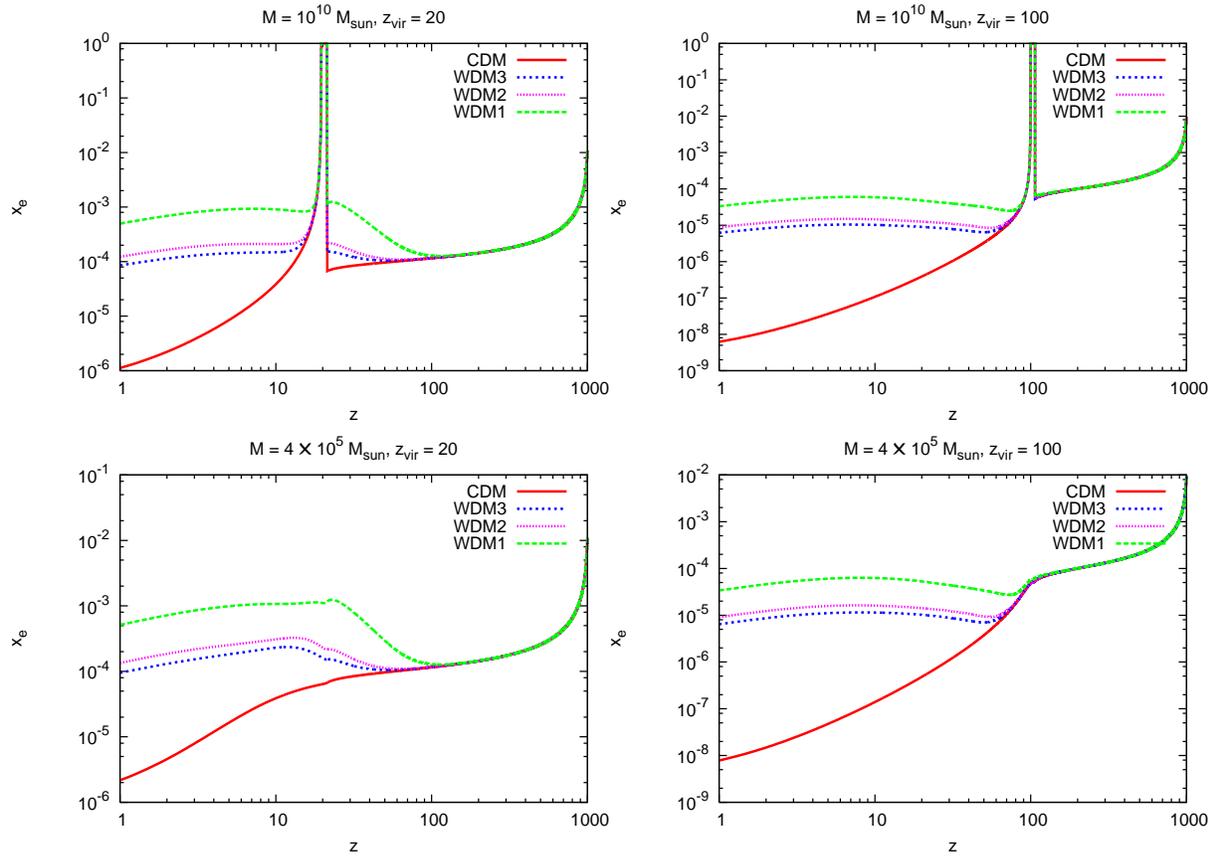}
\caption{The same as in Fig.~\ref{fig:T}, but for the ionization fraction. }
\label{fig:xe}
\end{figure}

\begin{figure}
\plotone{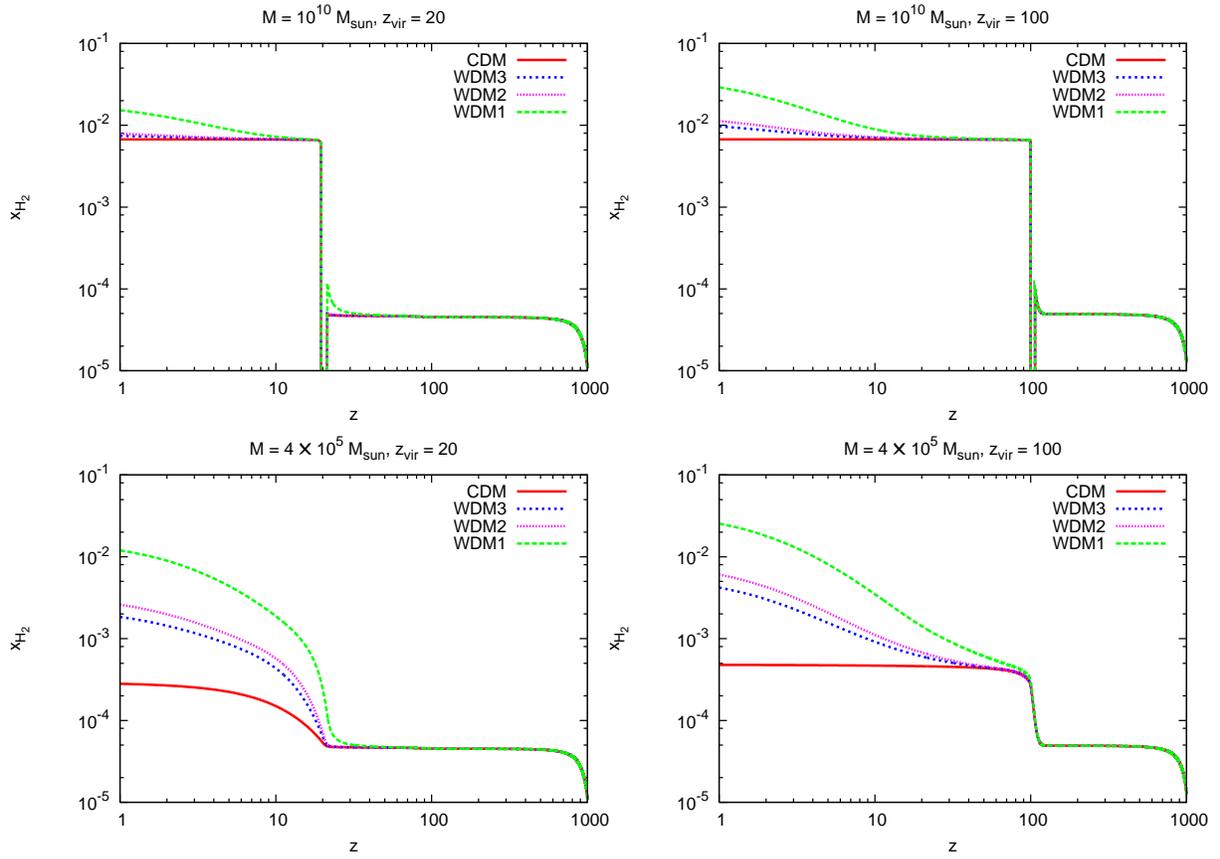}
\caption{The same as in Fig.~\ref{fig:T}, but for the molecular hydrogen
fraction.}
\label{fig:xH2}
\end{figure}

The results of the calculation are presented in Figures \ref{fig:T},
\ref{fig:xe} and \ref{fig:xH2}. Several
conclusions can be drawn from these results. We find that  large
clouds, with masses greater than $10^{10}M_\odot $, collapse regardless of the
details of cooling in all the models defined above. One could expect this
behavior because the virial temperature of such a cloud is higher than
$10^{4}$~K, so the H cooling mechanism is efficient enough and the additional
H$_{2}$ does not play an important role.

There is, however, a dramatic difference when it comes to
smaller clouds.  For a virialization redshift $z_{vir}=20$, a cloud of $4\times
10^5M_\odot $ collapses in model WDM1, but not in the CDM or other
WDM models.  This illustrates the importance of the increased H$_2$ fraction on
the cooling of gas.  

In WDM2 and WDM3 model, because of the relatively low associated x-ray flux, 
the temperature evolution is practically unaffected by the photons from the
dark matter decay.

From Fig.~\ref{fig:xH2} one can draw another conclusion. In the large
clouds, all of the H$_{2}$ molecules are destroyed during the virialization
process.  This is because of the very high virial temperature. In contrast,
there is a drastic increase in H$_{2}$ fraction during the virialization of 
the clouds with the mass of $4\times 10^5M_\odot $. In both cases, most of
molecular hydrogen is produced after virialization. 

For clouds with $M=4\times 10^5M_\odot $ and $z_{vir}=100$ the critical H$_{2}$
fraction (see eq. (11) of \citealt{tegmark97}) is reached at virialization. On
the other hand, in the clouds having equivalent mass with $z_{vir}=20$, the
molecular hydrogen faction $x_{H_{2}}$ attains its virial critical value only
for WDM1 model. This happens at $z=7.5$.

In all the WDM models the ionization and the molecular hydrogen fraction are 
enhanced in comparison with the CDM model, as suggested by \citet{biermann06}.
In addition, we can see that in WDM models the temperature of the collapsing
clouds soon after virialization is lower then that in the CDM case, which
speeds up the collapse, at least for the low-mass clouds.

We note in passing that the density of the cloud was kept constant after
virialization, so the evolution of $T$, $x_{e}$ and $x_{H_{2}}$ is reliable
for the redshifts close to virialization, but not for much smaller redshifts. 

It appears that WDM3 model is disfavored. The increase in the cooling rate 
is almost negligible in comparison with the CDM case, while the suppression on
the small scales is very strong. Therefore, this leads to an unacceptable
delay in the onset of star formation, as discussed by \citet{yoshida03}. 

In the case of WDM1 and WDM2 models, the cooling is enhanced by the additional
fraction of molecular hydrogen.  However, we cannot definitively prove that
star formation occurs at the redshift consistent with WMAP because of the
simplifying assumptions and limitations of our analysis.

The main limitation is that the virialization redshift of a
particular cloud is a fixed parameter in the simulation. In reality, it can be
changed by the enhancement of the cooling, so the cloud can reach its virial
stage much earlier.  Moreover, the density of the cloud is kept constant after
virialization, which is a simplifying assumption.  We can not
calculate the real redshift of the cloud collapse and onset of the star
formation. As a result, we cannot calculate how many clouds collapse at a 
given redshift and we cannot calculate the redshift of reionization in WDM
model reliably.  Also, we have not included the additional x-rays 
from the decay of sterile neutrinos inside the collapsing clump. This 
radiation can cause an additional increase in ionization and heating, 
which could change our results. Also the effects of opacity can
be important in large clouds, which we have not taken into account.
These limitations will be addressed in the future work.

\section{Conclusions}
\label{sec_conclusion}

We have performed a detailed analysis of the cooling and collapse of primordial
gas in the model with warm dark matter, taking into account both the increase
in the fraction of molecular hydrogen~\citep{biermann06} and the heating due to
the sterile neutrino decay.  As expected, the effect on the largest gas clouds
is negligible, but the smaller clouds are, in fact affected.  We have performed
the analysis for some benchmark cases which arise in  realistic scenarios.  For
the largest clouds, the additional molecular hydrogen made no difference.  For
smaller clouds, especially for those with masses $10^5-10^6~M_\odot$, the
increase in the x-ray background made the successful collapse possible in cases
where it could not occur in the absence of sterile neutrino decays.

\section*{Acknowledgments}

We thank G.~Gilmore, M.~Mapelli, M.~Shaposhnikov, and S.~Stachniewicz for very
helpful discussions and comments.   The work of P.L.B. and J.S. was supported
by the Pierre Auger grant 05~CU~5PD1/2 via DESY/BMBF. The work of A.K. was
supported in part by the DOE grant DE-FG03-91ER40662 and by the NASA ATP grants
NAG~5-10842 and NAG~5-13399.  A.K. thanks the CERN Theory unit and EPFL 
for hospitality during his visit.  J.S. thanks the organizers of  Marcel
Grossmann meeting for support. 
 
\section*{Appendix}
\appendix

\section{Specific intensity of the photons from the sterile neutrino decays}

In this section we shall calculate the photon spectrum produced by the
radiative
decays of the sterile neutrinos (our derivation is similar to the one in
\citet{masso99}). The dominant channel of radiative decays is through
the following reaction
\begin{equation}
	\nu_{s} \rightarrow \nu_{a} + \gamma \textrm{,}
\end{equation}
where $\nu_{s}$ is the sterile neutrino and $\nu_{a}$ is an ordinary active
neutrino. The photon energy is given by
\begin{equation}
	E_{0}=\frac{\left(m_{ s}^{2}+m_{ a}^{2}\right)}{2m_{ s}} \approx
\frac{m_{ s}}{2},
\end{equation}
where $m_{ s}$ and $m_{ a}$ are respectively the mass of the sterile and an
active neutrino. The cosmic expansion redshifts the photon energy. A photon
that at redshift $z$ has energy $E$ was produced at redshift $z_{0}$ given by
\begin{equation}
	1+z_{0}=\left(1+z\right) \frac{E_{0}}{E} \textrm{.} \label{energy}
\end{equation}

Let $F_{E}\left(z\right)$ be the energy flux (in units of erg $\rm{cm}^{-2}$
$\rm{s}^{-1}$) at redshift z of photons with energy E produced by the sterile
neutrino decays. The flux per unit energy and solid angle is given by
\begin{equation}
	\frac{d^{2} F_{E}\left(z\right)}{dE d \Omega}=E\frac{d^{2}
F_{n}\left(z\right)}{dE d \Omega} \textrm{,}
\end{equation}
where $F_{n}\left(z\right)$ is the photon flux at redshift z. It is related to
the photon flux at emission redshift $z_{0}$ through the following equation
\begin{equation}
  \frac{d^{2} F_{n}\left(z\right)}{d \Omega} =
\left(\frac{1+z}{1+z_{0}}\right)^3 \frac{d^{2} F_{n}\left(z_{0}\right)}{d
\Omega} = \frac{1}{4\pi} \left(\frac{1+z}{1+z_{0}}\right)^3 \delta n_{\gamma}
\left(z_{0}\right) c \textrm{,}
\end{equation}
where we have included the factor of $\left(\frac{1+z}{1+z_{0}}\right)^3$
produced by the expansion of the universe. The photon density emitted at
$z_{0}$ is given by the usual decay law
\begin{equation}
\delta	n_{\gamma} \left(z_{0}\right)= \frac{\delta t}{\tau} \widetilde{n}_{s}
\left(z_{0}\right)   \textrm{,}
\end{equation}
where $\tau$ is the mean sterile neutrino lifetime for the decay into an active
neutrino and a photon (see eq. \ref{meanlife}), $\widetilde{n}_{s}
\left(z_{0}\right)$ is the number density of the sterile neutrino at redshift
$z_{0}$ and $\delta t$ is given by
\begin{equation}
	\delta t =
\frac{dt}{dz_{0}}dz_{0}=-\frac{dz_{0}}{H_{0}\left(1+z_{0}\right)\sqrt{\Omega_{
\Lambda}+\Omega_{0}\left(1+z_{0}\right)^{3}}} 
= \frac{1}{H\left(z_{0}\right)} \frac{dE}{E} \textrm{,}
\end{equation}
where
$H\left(z_{0}\right)=H_{0}\sqrt{\Omega_{\Lambda}+\Omega_{0}\left(1+z_{0}
\right)^{3}}$ is the Hubble expansion rate at photon emission redshift $z_{0}$.
Writing everything together we obtain
\begin{equation}
	\frac{d^{2} F_{E}\left(z\right)}{dE d \Omega}=\frac{1}{4\pi}
\left(\frac{1+z}{1+z_{0}}\right)^3 \frac{\widetilde{n}_{s} \left(z_{0}\right)
c}{\tau H\left(z_{0}\right)} \textrm{.} \label{ff1}
\end{equation}

Choosing the time $t_{p}\ll \tau$ otherwise arbitrary we can write
\begin{equation}
\left(	\frac{1+z}{1+z_{0}}\right)^3  \widetilde{n}_{s} \left(z_{0}\right) =
\left(	\frac{1+z}{1+z_{p}}\right)^3 \widetilde{n}_{s} \left(z_{p}\right)
\exp \left(\frac{t_{p}-t\left(z_{0}\right)}{\tau}\right) = n_{s}\left(z\right)
\exp \left(\frac{t_{p}-t\left(z_{0}\right)}{\tau}\right) \label{neu}
\end{equation}
where $t\left(z_{0}\right)$ is the age of the universe at redshift $z_{0}$,
$t_{p}=t\left(z_{p}\right)$ and $n_{s}\left(z\right)$ would be the number
density of sterile neutrinos at redshift $z$ if they did not decay. 

Assuming that all of the dark matter consists of sterile neutrinos,
$n_{s}\left(z\right)$ can be written as
\begin{equation}
n_{s}\left(z\right) = n_{s}^{0} \left(1+z\right)^{3} \textrm{,}
\end{equation}
where
\begin{equation}
	n_{s}^{0} = \Omega_{dm}\frac{\rho_{0} }{m_{s}} \textrm{.}
\end{equation}

In the case of a flat universe filled with matter and a nonvanishing
cosmological constant the age of the universe at redshift $z$ is given by
\citep{masso99}
\begin{equation}
	t\left(z\right)=\frac{2}{3H_{0}\sqrt{\Omega_{\Lambda}}} \ln
\frac{\sqrt{\Omega_{\Lambda}}+\sqrt{\Omega_{\Lambda}+\Omega_{0}
\left(1+z\right)^3}}{\sqrt{\Omega_{0}\left(1+z\right)^3}} \textrm{.} 
\end{equation}

Taking $t_{p}=0$ and substituting eq. (\ref{neu}) into eq. (\ref{ff1}) we can
get the specific intensity (in units of erg $\rm{cm}^{-2}$ $\rm{s}^{-1}$
$\rm{sr}^{-1}$ $\rm{Hz}^{-1}$) of the photons with frequency $\nu$ from the
sterile neutrino decays at redshift $z$
\begin{equation}
	I_{\nu}\left(z\right)= h\frac{d^{2} F_{E}\left(z\right)}{dE d
\Omega}=\frac{h}{4\pi} \frac{n_{s} \left(z \right ) c}{\tau
H \left(z_{0}\right)}
\exp \left(\frac{-t\left(z_{0}\right)}{\tau}\right) \textrm{.} \label{ff2}
\end{equation}

We have to multiply this value by $e^{ -\tau_{abs} \left(z,z_{0} \right) }$ in
order to correct specific intensity for the absorption by the background
neutral hydrogen atoms.
The optical depth is given by
\begin{equation}
	\tau_{abs}\left(z,z_{0}\right) =  \int_{z}^{z_{0}} 
\sigma_{12}\left(E\left(z'\right)\right) n_{H}\left(z'\right) c 
\frac{dt}{dz'} 
dz' \textrm{,} \label{abs}
\end{equation}
where $\sigma_{12}\left(E\left(z'\right)\right)$ and $n_{H}\left(z'\right)$ are
the cross section for the H ionization at energy $E\left(z'\right)$ and the
hydrogen number density of the Universe at redshift $z'$ respectively;
$E\left(z'\right)$ is the energy of the photon at redshift $z'$, which was
emitted in sterile neutrino decay at redshift $z_{0}$ and is given by the
equation
\begin{equation}
	E\left(z'\right)=\frac{1+z'}{1+z_{0}}E_{0} \textrm{.} \label{ap1}
\end{equation}

The mean hydrogen number density of the Universe at redshift $z'$ can be
written as
\begin{equation}
	n_{H}\left(z'\right)=n^{0}_{H}\left[1-x_{e}\left(z\right)\right]
\left(1+z'\right)^3 \textrm{,} \label{ap2}
\end{equation}
where $n^{0}_{H}$ is its present value and is equal to
\begin{equation}
	n^{0}_{H}=\frac{\Omega_{b}\rho_{0}X}{m_{H}} \textrm{.} \label{ap3}
\end{equation}

The cross section for the H ionization is given by
\begin{equation}
	\sigma_{12}\left(E\left(z'\right)\right)=\sigma_{0}\left(\frac{h
\nu^{H}_{th}}{E\left(z'\right)}\right)^3 \textrm{,} \label{ap4}
\end{equation}
where $\sigma_{0}=7.909 \times 10^{-18}$ cm$^{-2}$ and $h \nu^{H}_{th}=13.6$
eV.

Neglecting the $\Omega_{\Lambda}$ term in eq. (\ref{hubble}), which is a very
good approximation at high redshifts, and using eqs.~(\ref{energy}),
(\ref{ap1}), (\ref{ap2}) and (\ref{ap4}) we can cast eq.
(\ref{abs}) into the form
\begin{equation}
	\tau_{abs}\left(z,z_{0}\right) =   \sigma_{0}
\left(\frac{h\nu_{th}}{E}\right)^3 \left(1+z\right)^3 n_{H}^{0}c
H_{0}^{-1} \Omega_{0}^{-1/2} \int_{z}^{z_{0}}
\left[1-x_{e}\left(z'\right)\right]\left(1+z'\right)^{-5/2}  dz' \textrm{.}
\end{equation}
As $x_{e}\left(z\right)$, let us take 
\begin{equation}
	x_{e} \left(z\right)=
	\left\{
	\begin{array}{ll}
	    0  \hskip 0.5cm z < z_{rec} & \\
	    1 \hskip 0.5cm z > z_{rec} & \textrm{,}
	\end{array}
	\right.
\end{equation}
where $z_{rec}$ is the recombination redshift. It gives us
\begin{equation}
	\tau_{abs}\left(z,z_{0}\right) =  \frac{2}{3} \sigma_{0}
\left(\frac{\nu_{th}}{\nu}\right)^3 \left(1+z\right)^{3/2} n_{H}^{0}c
H_{0}^{-1} \Omega_{0}^{-1/2} \left[
1-\left(\frac{1+z}{1+\textrm{min}\left(z_{0},z_{rec}\right)}\right)^{3/2}
\right] \textrm{.}
\end{equation}

Photons emitted before recombination are Compton scattered and thermalized. For
simplicity we assume that these photons are not visible. The final specific
intensity is equal to
\begin{equation}  I_{\nu}\left(z\right)= 
\left\{
	\begin{array}{ll}
	    \frac{	h c n_{s}\left(z\right) e^{
-\frac{t\left(z_{0}\right)}{\tau}} e^{ -\tau_{abs} \left(z,z_{0} \right) }}{4
\pi \tau H_{0}
\sqrt{\Omega_{\Lambda}+\Omega_{0}\left(1+z\right)^{3}\left(\frac{E_{0}}{h
\nu}\right)^3} } & \hskip 0.0cm h\nu > E_{0} \frac{1+z}{1+z_{rec}}  \\
	    0 & \hskip 0.0cm h\nu < E_{0} \frac{1+z}{1+z_{rec}}  
	\end{array} 
	\right.
 \textrm{,}  \label{f1}
\end{equation}
where
\begin{equation}
t\left(z_{0}\right) = t\left(\frac{E_{0}}{h \nu}\left(1+z\right)-1\right)
\textrm{,} 
\end{equation}
\begin{equation}
	\tau_{abs} \left(z,z_{0} \right)= \tau_{abs} \left(z,\frac{E_{0}}{h
\nu}\left(1+z\right)-1 \right) \textrm{.}
\end{equation}

We have assumed that the photons are absorbed only by the Ly$\alpha$ wedge, so
the above equations are valid only for $h\nu>13.6$ eV. To derive the photon
flux at energy lower than 13.6 eV we have to take into account another hydrogen
absorption lines. 
The radiation with energy below 13.6 eV start to be important only at lower
redshifts (e.g. for $m_{s}=5$ keV at $z\approx5$) so it is not relevant in our 
calculations.

\section{Heating and ionization due to the sterile neutrino decays}

The photons from the decays of the sterile neutrino are mainly absorbed by
neutral hydrogen atoms leading to their ionization. The ionization rate due to
these photons is enhance almost 100 times due to additional ionization by the
secondary electrons which deposit almost 1/3 of its energy into ionization
(depending on the ionization fraction of the medium). The
energy of the absorbed photons partially goes into ionization and partially
into heating and excitations. We have adopted the approximation by
\citet{steen}, for which ionization rate and heating due to the photons from
the decays of the sterile neutrino are respectively equal to
\begin{eqnarray}
A\left(z\right) &=& \left[ \int_{\nu^{H}_{th}}^{\infty}4\pi  
\sigma_{H}\left(\nu \right)  
\frac{I_{\nu}\left(z\right)}{h\nu} \left(\frac{h \nu -h \nu_{th}^{H}}{h
\nu_{th}^{H}} \right) d\nu \right. \nonumber \\
&+& \left. \frac{Y}{4X} \int_{\nu^{He}_{th}}^{\infty}4\pi  
\sigma_{He}\left(\nu \right)  
\frac{I_{\nu}\left(z\right)}{h\nu} \left(\frac{h \nu -h \nu_{th}^{He}}{h
\nu_{th}^{H}} \right) d\nu \right] \nonumber \\
&\times& C_{i}\left(1-x_{e}^{a_{i}}\right)^b_{i} +
\int_{\nu_{th}^{H}}^{\infty}4\pi\sigma_{H}\left(\nu\right)\frac{I_{\nu}
\left(z\right)}{h\nu} d\nu \label{ion}
\end{eqnarray}
\begin{eqnarray}
\Gamma_{s}\left(z\right) &=&  \left[ \int_{\nu^{H}_{th}}^{\infty}4\pi  
\sigma_{H}\left(\nu \right)  
\frac{I_{\nu}\left(z\right)}{h\nu} \left(h \nu -h \nu_{th}^{H}  \right) d\nu
\right. \nonumber \\
&+& \left. \frac{Y}{4X} \int_{\nu^{He}_{th}}^{\infty}4\pi  
\sigma_{He}\left(\nu \right)  
\frac{I_{\nu}\left(z\right)}{h\nu} \left(h \nu -h \nu_{th}^{He}  \right) d\nu
\right] \nonumber \\
&\times& C_{h}\left[ 1-\left(1-x_{e}^{a_{h}}\right)^b_{h} \right] n_{H}
\label{heat} \textrm{,}
\end{eqnarray}
where
\begin{equation}
	\sigma_{He}\left(\nu \right) = 7.83 \times 10^{-18} \textrm{cm}^{-2}
\left[  1.66\left(\frac{ \nu_{th}^{He}}{\nu }\right)^{2.05} - 0.66\left(
\frac{\nu_{th}^{He}}{\nu} \right)^{3.05}  \right] 
\end{equation}
is the cross section for He ionization, $h \nu^{He}_{th}=24.6$~eV is its
energy threshold;  $C_{i}=0.3908$, $a_{i}=0.4092$,
$b_{i}=1.7592$, $C_{h}=0.9971$, $a_{h}=0.2663$, $b_{h}=1.3163$ \citep{steen}. 

\section{Cooling/heating function}
\label{sec_cooling_heating}

The H$_{2}$ cooling function ($\Lambda_{H-H_{2}}$, $\Lambda_{H_{2}-H_{2}}$ and
$\Lambda_{form}$) given in Table \ref{tab:cooling} has been computed assuming
that the CMB temperature is much smaller than the gas temperature. This
approximation is not valid in our case since in the WDM model with keV sterile
neutrinos the gas temperature can be very close or even lower than the CMB
temperature because of relatively high ionization fraction. The situation is
very similar in CDM model, where the gas temperature strictly follows the CMB
temperature at least at high redshifts.  

In such conditions, the level population of H$_{2}$ is strongly affected by
stimulated emission and absorption. Molecules become an effective heating
source for the gas, because the rate of collisional de-excitation of the 
vibrational levels is faster than their radiative decay. In order to
correct the cooling/heating function for this effect, we have to take the net
cooling rate
\begin{equation}
	\Lambda_{i} \left( T, T_{r}\right) = \Lambda_{i} \left( T\right) -
\Lambda_{i} \left( T_{r}\right) \textrm{,}
\end{equation}
where $\Lambda_i$ represents either $\Lambda_{H-H_{2}}$,
$\Lambda_{H_{2}-H_{2}}$, or $\Lambda_{form}$.  We also note that
$\Lambda_{i} \left( T, T_{r}\right) \approx  \Lambda_{i} \left( T\right)$ for
$T \gg T_{r}$. 

One should do the same for the $\Lambda_{H-e}$ (see Table
\ref{tab:cooling}). However, the cooling from the collisional excitation of
hydrogen atoms H and its subsequent spontaneous de-excitation is only
important for the gas temperatures $T>10^{4}$. In our case the CMB temperature
is well below $10^4$ K, so we can neglect the heating term for this
cooling/heating mechanism.

For completeness we have to take also into account the fact that electrons can
gain energy from the CMB radiation when $T<T_{r}$, so we are taking as the
bremsstrahlung cooling/heating term
\begin{equation}
	\Lambda_{brem}\left(T,T_{r}\right)=\Lambda_{brem}
\left(T\right)-\Lambda_{brem}\left(T_{r}\right) \textrm{.}
\end{equation}

In our calculations we have used the following cooling/heating function: 
\begin{eqnarray}
	\Lambda \left(T,T_{r},z\right)
&=&
\Lambda_{H}\left(T,T_{r}\right)+\Lambda_{H_{2}}\left(T,T_{r}\right)+\Lambda_
{comp}\left(T,T_{r}\right) \nonumber \\
	&+& \Lambda_{brem}\left(T,T_{r}\right)+\Lambda_{s}\left(z\right) 
\textrm{.} \label{coolfunc}
\end{eqnarray}

\clearpage

\tabletypesize{\scriptsize}
\begin{deluxetable}{lll}
\tablecolumns{3}
\tablewidth{0pc}
\tablecaption{Reaction rates}
\tablehead{
\colhead{Reaction} & \colhead{Rate ($\rm{cm}^{3}$
$\rm{s}^{-1}$ or $\rm{s}^{-1}$)} & \colhead{Reference} }
\startdata
\multicolumn{3}{c}{H$_{2}$ formation via $H^{-}$ channel} \\
\\
   $e^{-}+H\rightarrow H^{-}+\gamma$ &
$k_{1}=1.4\times10^{-18}T^{0.928}\exp\left(-\frac{T}{16200}\right)$ &
\citep{galli}	\\
   $H+H^{-}\rightarrow H_{2}+e^{-}$	 & $k_{2}=4.0\times10^{-9}T^{-0.17}$
&	\citep{galli}		\\
   $H^{-}+\gamma \rightarrow H+e^{-}$	  & $k_{3}=1.1\times10^{-1}
T_{r}^{2,13}\exp\left(-\frac{8823}{T_{r}}\right)$ & \citep{galli} \\
 \\
   \multicolumn{3}{c}{H$_{2}$ formation via H$_{2}^{+}$ channel} \\
   \\
   $H^{+}+H\rightarrow H_{2}^{+}+\gamma$& $\log k_{4}=-19.38 -1.523\log T
+1.118\left(\log T\right)^{2} $ & \\ &  $-0.1269\left(\log T\right)^{3}$
& \citep{galli} \\
   H$_{2}^{+}+H\rightarrow H_{2}+H^{+}$	& $k_{5}=6.4\times10^{-10}$ &
\citep{galli}	\\   
   H$_{2}^{+}+\gamma \rightarrow H+H^{+}$	&
$k_{6}=2.0\times10^{1}T_{r}^{1.59}\exp\left(-\frac{82000}{T_{r}}\right)$
& \citep{galli}	\\   
   \\
   \multicolumn{3}{c}{Collisional dissociation of H$_{2}$} \\
   \\ 
   H$_{2}+H\rightarrow 3H$ & $k_{7}=k_{7,H}^{1-a}k_{7,L}^{a}$  \\	
	 &
$k_{7,L}=1.12\times10^{-10}\exp\left(-\frac{7.035\times10^4}{T}\right)$
& \citep{shapiro} \\	
&$k_{7,H}=6.5\times10^{-7}T^{-1/2}\exp\left(-\frac{\chi_{H_{2}}}{kT}
\right)\left
[1-\exp\left(-\frac{6000}{T}\right)\right]$ & \citep{palla83} \\   									
		 & $a=\left(1+\frac{n_{H}}{n_{cr}}\right)^{-1}$ \\   							
	 & $\log n_{cr}=4.0-0.416\log T_{4}-0.327\left(\log T_{4}\right)^{2}$
& \citep{shapiro} \\
   H$_{2}+H_{2}\rightarrow 2H+H_{2}$ & $k_{8}=k_{8,H}^{1-a}k_{8,L}^{a}$ &
\citep{shapiro} \\   										  
& $k_{8,L}=1.18\times10^{-10}\exp\left(-\frac{6.95\times10^{4}}{T}\right)$
\\					  
&$k_{8,H}=1.3\times10^{-9}\exp\left(-\frac{5.33\times10^{4}}{T}\right)$ \\   								
			& $a=\left(1+\frac{n_{H_{2}}}{n_{cr}}\right)^{-1}$ \\	
						 & $\log n_{cr}=4.845-1.3\log
T_{4}+1.62\left(\log T_{4}\right)^{2}$ \\   									 
   H$_{2}+e^{-}\rightarrow 2H+e^{-}$ &
$k_{9}=1.3\times10^{-18}T^{2}\exp\left(-\frac{\chi_{H_{2}}}{kT}\right)$ &
\citep{lenzuni} \\
   
   \\
   \multicolumn{3}{c}{Photo-dissociation of H$_{2}$} \\
   \\
   
   H$_{2}+\gamma \rightarrow H_{2}^{*}\rightarrow 2H$ &
$k_{10}=1.1\times10^{8}j\left(\bar{\nu}\right)$ & \citep{abel} \\ 
		& $j\left(\bar{\nu}\right)$ is the CMB
radiation
flux in erg $\rm{cm}^{-2}$ $\rm{s}^{-1}$ \\ & at $h\bar{\nu}=12.87 \textrm{
eV}$ & \\
   H$_{2}+\gamma \rightarrow 2H$ &
$k_{11}=\int_{\nu^{p}_{th}}^{\infty}4\pi\sigma_{11}\left(\nu\right)\frac{J_{\nu
}}{h\nu}d\nu$ & \citep{abel}\\
& $h\nu^{p}_{th}=14.159$ eV, \\ & 
$J_{\nu}=\frac{2h\nu^{3}}{c^2} \frac{1}{\exp\left(\frac{h\nu}{kT_{r}}
\right)-1}$ is the CMB intensity  & \\
& $\sigma_{11}\left(\nu\right)=\frac{1}{1+y}\sigma^{p}_{11}\left(\nu\right)+\frac
{
y}{1+y}\sigma^{o}_{11}\left(\nu\right)$ \\                                
                                 & $\sigma^{p}_{11}$ and $\sigma^{o}_{11}$ are
the crossections for para- and ortho-H$_{2}$  \\
                                 & respectively, $y \approx 3$ is the ortho-,
to para-H$_{2}$ ratio \\
    
   \\
   \multicolumn{3}{c}{Ionization and recombination of H} \\
   \\
   
   $H+\gamma \rightarrow H^{+}+e^{-}$ &
$k_{12}=\int_{\nu^{H}_{th}}^{\infty}4\pi\sigma_{12}\left(\nu\right)\frac{J_{\nu
}}{
h\nu}d\nu$ & \citep{rybicki} \\  
                                      & $h\nu^{H}_{th}=13.6$ eV \\
                                      &
$\sigma_{12}=7.909\times10^{-18}\textrm{
cm}^{-2}\left(\frac{\nu}{\nu_{th}}\right)^{-3}$ \\
   $H^{+}+e^{-} \rightarrow H+\gamma$ &
$k_{13}=8.4\times10^{-11}T^{-1/2}T_{3}^{-0.2}\left(1+T_{6}^{0.7}\right)^{-1}$ &
\citep{galli}  \\       
   $H+e^{-} \rightarrow H^{+}+2e^{-}$ &
$k_{14}=5.85\times10^{-11}T^{1/2}\left(1+T_{5}^{1/2}\right)^{-1}\exp
\left(-\frac{157809.1}{T}\right)$ &
\citep{haiman96} \\      
      \\
   \multicolumn{3}{c}{Ionization of H by the photons from the sterile
neutrino decays} \\
    \multicolumn{3}{c}{(Ly$\alpha$ ionization wedge and ionization by the
secondary electrons)} \\
   \\      
      $H+\gamma \rightarrow H^{+}+e^{-}$ & see text eq. (\ref{ion}) &  \\
\enddata
\tablecomments{
The binding energy of H$_{2}$ is $\chi_{H_{2}} / k=5.197\times10^{4}$~K.
$T$ and $T_{r}=2.725\left(1+z\right)$ are the gas and the CMB temperatures, 
respectively.  $J_{\nu}$ is the CMB intensity. $I_{\nu}\left(z\right)$ is the
intensity of the photons from the sterile neutrino decays at redshift $z$.
$T_{n}=T/10^{n}$ K. $x_{e}\left(z\right)$, $z_{rec}$ and $E_{0}$ are the
ionization fraction at redshift $z$, redshift of recombination and the energy
of the photon produced in the sterile neutrino decays.}
\label{tab:reaction}
\end{deluxetable}

\tabletypesize{\scriptsize}
\begin{deluxetable}{ll}
\tablecolumns{2}
\tablewidth{0pc}
\tablecaption{Cooling/heating processes}
\tablehead{
\colhead{Cooling/heating function (erg $\rm{cm}^{-3}$
$\rm{s}^{-1}$)}  & \colhead{Reference} }
\startdata 
  \multicolumn{2}{c}{H cooling}  \\
  \\
  \multicolumn{2}{c}{$ \Lambda_{H}\left( T,
T_{r}\right)=\Lambda_{H-e}\left(T\right) + \Lambda_{rec}\left(T\right) +
	\Lambda_{ioniz}\left(T\right)-
	\Gamma_{PD}\left(T_{r}\right) $}\\
  \\
   a) excitation of H by collisions with $e^{-}$\\
   
$\Lambda_{H-e}\left(T\right)=7.5\times10^{-19}\left(1+T_{5}^{1/2}
\right)^{-1} \exp \left(-\frac{118348}{T}\right) n_{e}n_{H}$ &
\citep{haiman96} 
\\
   \\
   b) recombination to H \\
   
    $\Lambda_{rec}\left(T\right)=1.04\times10^{-16}k_{13}n_{e}n_{H}^{+}$ &
\citep{haiman96}\\
    \\
   c) ionization of H by collisions with $e^{-}$\\
    $ \Lambda_{ioniz}\left(T\right)=2.17\times10^{-11}k_{14}n_{e}n_{H}$ &
\citep{haiman96} \\
   \\
    d) photo-ionization of H
    \\
     $ \Gamma_{PD}\left(T_{r}\right)=4\pi n_{H}
\int_{\nu^{H}_{th}}^{\infty}\sigma_{12}\left(\nu\right)\frac{J_{\nu}}{h\nu}
\left(  h
\nu - h\nu^{H}_{th}\right)d\nu$  \\  
      $ h\nu^{H}_{th}=13.6$ eV \\    
   
   \\
   \multicolumn{2}{c}{H$_{2}$ cooling} \\
   \\
   
   \multicolumn{2}{c}{$ \Lambda_{H_{2}}\left( T,
T_{r}\right)=\Lambda_{H-H_{2}}\left(T,T_{r}\right) + 
	\Lambda_{H_{2}-H_{2}}\left(T,T_{r}\right) +
	\Lambda_{form}\left(T,T_{r}\right)+
	\Lambda_{diss}\left(T\right)-\Gamma_{PD1}\left(T_{r}\right)-
	\Gamma_{PD2}\left(T_{r}\right) $}\\ 
   \\
a) collisional excitation of H$_{2}$ \\
\\
\multicolumn{2}{l}{i) excitation by collisions with H $\hspace{1 cm}
\Lambda_{H-H_{2}}\left(T,T_{r}\right)=\Lambda_{H-H_{2}}\left(T\right)
-\Lambda_{H-H_{2}}\left(T_{r}\right)$} \\
 $ \Lambda_{H-H_{2}}\left(T\right)=
\frac{\Lambda^{H}\left(LTE\right)\Lambda^{H}\left(n\rightarrow0\right)}{
\Lambda^
{H}\left(n\rightarrow0\right)+\Lambda^{H}\left(LTE\right)}n_{H_{2}}n_{H}$ &
\citep{galli} \\
 \\
 \multicolumn{2}{l}{ii) excitation by collisions with H$_{2}$ $\hspace{1
cm} \Lambda_{H_{2}-H_{2}}\left(T,T_{r}\right)=\Lambda_{H_{2}-H_{2}}
\left(T\right) -\Lambda_{H_{2}-H_{2}}\left(T_{r}\right)$} \\
\\
  $ \Lambda_{H_{2}-H_{2}}\left(T\right)=\left( 
\frac{\Lambda_{r}^{H_{2}}\left(LTE\right)\Lambda_{r}^{H_{2}}
\left(n\rightarrow0\right)}{\Lambda_{r}^{H_{2}}
\left(n\rightarrow0\right)+\Lambda_{r}^{H_{2}}\left(LTE\right)}+
\frac{\Lambda_{v}^{H_{2}}\left(LTE\right)\Lambda_{v}^{H_{2}}
\left(n\rightarrow0\right)}
{\Lambda_{v}^{H_{2}}\left(n\rightarrow0\right)+\Lambda_{v}^{H_{2}}
\left(LTE\right)}
\right)n_{H_{2}}^{2}$ & \citep{hollenbach} \\  
\\
   $
\Lambda^{H}\left(LTE\right)=\Lambda_{r}^{H}\left(LTE\right)+\Lambda_{v}^{H}
\left(LTE\right)$ & \citep{hollenbach} \\
  $ \log \Lambda^{H}\left(n\rightarrow0\right)=-103+97.59 \log T -48.05
\left(\log T\right)^{2}+10.8\left(\log T\right)^{3}-0.9032\left(\log
T\right)^{4}$ & \citep{galli} \\
   $
\Lambda_{r}^{H,H_{2}}\left(LTE\right)=\frac{1}{n_{H,H_{2}}}\left\{\left(\frac{
9.5\times10^{-22}T_{3}^{3.76}}{1+0.12T_{3}^{2.1}}\right)\exp
\left[-\left(\frac{0.13}{T_{3}}\right)^{3}\right]+
  3\times10^{-24}\exp\left(-\frac{0.51}{T_{3}}\right)\right\}$ &
\citep{hollenbach} \\   
    $
\Lambda_{v}^{H,H_{2}}\left(LTE\right)=\frac{1}{n_{H,H_{2}}}\left[6.7\times10^{
-19}\exp\left(-\frac{5.86}{T_{3}}\right)+1.6\times10^{-18}\exp\left(-\frac{11.7
}
{T_{3}}\right)\right]$ & \citep{hollenbach} \\
  
 $
\Lambda_{r}^{H_{2}}\left(n\rightarrow0\right)=0.25\left[5\gamma^{H_{2}}_{2}
\exp\left(-\frac{E^{r}_{20}}{kT}\right)E^{r}_{20}\right]+0.75\left[\frac{7}{3}
\gamma_{3}^{H_{2}}\exp\left[-\left(\frac{E^{r}_{31}}{kT}\right)\right]E^{r}_{31
}
\right]$ & \citep{hollenbach} \\
 
 $
\Lambda_{v}^{H_{2}}\left(n\rightarrow0\right)=1.4\times10^{-12}T^{1/2}
\exp\left(-\frac{12000}{T+1200}\right) 
 \exp\left(-\frac{E^{v}_{10}}{kT}\right)E^{v}_{10}$ & \citep{hollenbach}\\
 $
\gamma_{J}^{H_{2}}=\left(3.3\times10^{-12}+6.6\times10^{-12}T_{3}\right)\left\{
0.276J^{2}\exp\left[-\left(\frac{J}{3.18}\right)^{1.7}\right]\right\}$ &
\citep{hollenbach} \\ 
 
 $ \frac{3}{5} E^{r}_{31}/k=E^{r}_{20}/k=512$ K, $E^{v}_{10}/k=5860$ K &
\citep{hollenbach} \\
 
  \\
  b) H$_{2}$ formation cooling $\hspace{1 cm}
\Lambda_{form}\left(T,T_{r}\right)=\Lambda_{form}\left(T\right) -\Lambda_{form}
\left(T_{r}\right)$\\
     $
\Lambda_{form}\left(T\right)=1.6022\times10^{-12}\left(3.53\frac{k_{1}k_{2}}{k_
{
2}+k_{3}/n_{H}}+1.83\frac{k_{4}k_{5}}{k_{5}+k_{6}/n_{H}}\right)\left[
1-\left(1+\frac{n_{cr}}{n}\right)^{-1}\right]n_{H}n_{e}$  & \citep{shapiro}  \\
 $
n_{cr}=10^{6}T^{-1/2}/\left\{1.6x_{H}\exp\left[-\left(\frac{400}{T}\right)^{2}
\right]+1.4x_{H_{2}}\exp\left[-\frac{12000}{T+1200}\right]\right\}$ &
\citep{hollenbach} \\
 
 \\
c) collisional dissociation of H$_{2}$ \\
$\Lambda_{diss}\left(T\right)=7.17145\times10^{-12}\left(n_{H}k_{7}+n_{H_{2}}k_{8}+n_{e}k_{9}\right)n_{H_{2}}$ & \citep{shapiro} \\
   \\
d) photo-dissociation of H$_{2}$ \\
\\
\multicolumn{2}{l}{i) H$_{2}+\gamma\rightarrow H_{2}^{*}\rightarrow2H$} \\

 $ \Gamma_{PD1}\left(T_{r}\right)=6.4\times10^{-13} k_{10}n_{H_{2}}$ &
\citep{abel} \\
\\
\multicolumn{2}{l}{ii) H$_{2}+\gamma\rightarrow2H$} \\
$ \Gamma_{PD2}\left(T_{r}\right)=4 \pi n_{H_{2}} \left[

 \frac{1}{y+1}\int_{\nu^{p}_{th}}^{\infty} \sigma^{p}_{11}\left(\nu\right)
\frac{J_{\nu}}{h\nu}\left(h\nu-h\nu^{p}_{th}\right)d \nu
 +
 \frac{y}{y+1}\int_{\nu^{o}_{th}}^{\infty} \sigma^{o}_{11}\left(\nu\right)
\frac{J_{\nu}}{h\nu}\left(h\nu-h\nu^{o}_{th}\right)d \nu
 \right]$ & \citep{abel}\\
$ h\nu^{p}_{th}=14.159$ eV, $h\nu_{th}^{o}=14.675$ eV, $y=3$ is the
ortho-H$_{2}$ to para-H$_{2}$ ratio \\

  \\
  \multicolumn{2}{c}{Compton cooling} \\

$\Lambda_{comp}\left(T,T_{r}\right)=1.017\times10^{-37}T_{r}^{4}
\left(T-T_{r} \right)n_{e}$ & \citep{haiman96} \\
  \\ 
  \multicolumn{2}{c}{bremsstrahlung} \\
  \\
 
\multicolumn{2}{c}{$\Lambda_{brem}\left(T,T_{r}\right)=\Lambda_{brem}
\left(T\right)-\Lambda_{brem}\left(T_{r}\right)$}\\
  \\
   
$\Lambda_{brem}\left(T\right)=1.42\times10^{-27}g_{ff}T^{1/2}n_{e}n_{
H^{+}}$  & \citep{haiman96} \\
	$ g_{ff}=1.1+0.34\exp\left[-\left(5.5-\log
T\right)^{2}/3\right]$ \\   
  \\
   \multicolumn{2}{c}{heating due to the photons from the sterile neutrino
decays} \\
    \\
   \multicolumn{2}{c}{$\Lambda_{s} \left(z\right) = -\Gamma_{s}\left(z\right)$}
\\
    \\  
    $$see text eq. (\ref{heat}) &  \\
  
  \enddata
\tablecomments{
$T$ and $T_{r}=2.725\left(1+z\right)$ are the gas and CMB temperature
respectively. $J_{\nu}$ is the CMB intensity. $I_{\nu}\left(z\right)$ is the
intensity of the photons from the sterile neutrino decays at redshift $z$.
$T_{n}=T/10^{n}$ K. $\sigma_{i}\left(\nu\right)$ and $k_{i}$ are given in the
table \ref{tab:reaction}. $x_{e}\left(z\right)$, $z_{rec}$ and $E_{0}$ are the
ionization fraction at redshift $z$, redshift of recombination and the energy
of the photon produced in the sterile neutrino decays.
}
\label{tab:cooling}
\end{deluxetable}


\end{document}